\documentclass[aps,prx,showpacs,amsmath,amssymb,amsfonts,superscriptaddress,twocolumn,longbibliography]{revtex4-1}
\usepackage{graphicx}
\usepackage{subfigure}
\usepackage{verbatim}
\usepackage{dcolumn}
\usepackage{bm}
\usepackage{color}
\usepackage[colorlinks=true,citecolor=blue,linkcolor=blue]{hyperref}
\usepackage{epstopdf}
\usepackage{soul}
\sethlcolor{yellow}

\newcommand{\bla}{bla\\bla\\bla\\bla\\bla}

\begin{document}

\title{Design principles and optimal performance for molecular motors under realistic constraints}

\author{Yuhai Tu}
\affiliation{IBM T.J. Watson Research Center, PO Box 218, Yorktown Heights, NY~10598, U.S.A.}
\email{yuhai@us.ibm.com}
\author{Yuansheng Cao}
\affiliation{Department of Physics, UCSD, La Jolla, CA92093}

\begin{abstract}



The performance of a molecular motor, characterized by its power output and energy efficiency, is investigated in the motor design space spanned by the stepping rate function and the motor-track interaction potential. Analytic results and simulations show that a gating mechanism that restricts forward stepping in a narrow window in configuration space is needed for generating high power at physiologically relevant loads. By deriving general thermodynamics laws for nonequilibrium motors, we find that the maximum torque (force) at stall is less than its theoretical limit for any realistic motor-track interactions due to speed fluctuations. Our study reveals a tradeoff for the motor-track interaction: while a strong interaction generates a high power output for forward steps, it also leads to a higher probability of wasteful spontaneous back steps. Our analysis and simulations show that this tradeoff sets a fundamental limit to the maximum motor efficiency in the presence of spontaneous back steps, i.e., loose-coupling. Balancing this tradeoff leads to an optimal design of the motor-track interaction for achieving a maximum efficiency close to $1$ for realistic motors that are not perfectly coupled with the energy source.
Comparison with existing data and suggestions for future experiments are discussed.

\end{abstract}


\maketitle
\newpage


\section{Introduction}
Molecular motors are essential for living systems. They convert chemical energy to mechanical work driving motion and transport in biological systems. 
While linear motors such as kinesin and myosin are fueled by ATP, bacterial flagellar motor (BFM) couples ion (e.g., $H^+$ and $Na^+$) translocations across cytoplasmic membrane to the rotation of flagellar filaments which propel the bacterial motion (tumbling or swimming) \cite{HBerg1973,Larsen1974,Hirota1981,HBerg2003}. A fundamental question is whether there are thermodynamic bounds to the power generation and energy efficiency for these highly non-equilibrium molecular engines \cite{Parmeggiani1999,Parrondo2002,Astumian2010}. A related and perhaps more important question is what are the microscopic properties (design features) that would allow a molecular motor to approach these bounds under realistic constraints. Here, we try to address these general questions and test the findings in the specific case of BFM, which is believed to be highly efficient. 

We first describe briefly what is known about the bacterial flagellar motor  (see \cite{Minamino2014} for a recent review). The rotor of this nanoscale rotary engine contains a ring of $\sim26$ FliG proteins (see \cite{Lee2016} for an alternative view of $34$ FliGs in the rotor), which serve as the track of the engine and interact with multiple torque-generating stator units that are anchored to the cell wall. In \emph{E. coli}, each stator unit is composed of four copies of MotA and two copies of MotB, forming two transmembrane proton channels \cite{Asai1997,Blair1990,Sato2000,Kojima2004,Yorimitsu2004,Chun1988,Roujeinikova2008}. Ion translocations through the channels cause conformational changes of the stator proteins which generate torque on the rotor to drive its rotation \cite{Block1984,Blair1988}. The ion flow is powered by the ion motive force (IMF), which is the free energy difference of an ion across the cell membrane. IMF depends on the transmembrane voltage and the ion concentration difference across the cytoplasmic membrane. For \emph{E. coli}, the responsible ion is proton, and the driving force is the proton motive force (PMF).

The mechanical properties of the flagellar motor, characterized by its torque-speed relationship, have been measured experimentally under various conditions (e.g., different PMF, temperature, number of stators) \cite{Manson1980,Khan1983,Lowe1987,Chen2000}. For \emph{E. coli}, the torque-speed dependence for a BFM in the counterclockwise (CCW) rotational state has a concave down shape, with a plateau of high torque at low speeds and a rapid drop of torque at high speeds. On the other hand, the torque-speed curve for the clockwise (CW) motor is almost linear \cite{Yuan2010}. Based on specific choices of the stator-rotor interaction and the energy transduction process, several models have been developed to explain the observed torque-speed relationship for the BFM \cite{Lauger1988,Berry1993,Xing2006,Mora2009,Meacci2009,Van2009,Meacci2011,Boschert2015,Mandadapu2015}.

Our understanding of the thermodynamics and energetics of BFM remains limited. Some experiments suggested that BFM is tightly coupled, meaning that a fixed number of ions pass through the motor per revolution \cite{Meister1987,Nakamura2010}. It was argued that since at high loads the motor moves slowly and thus operates near equilibrium with the thermal bath, the efficiency should be close to one \cite{Meister1989}. However, recent experiments by Lo \emph{et al.} \cite{Lo2013} found that the maximum torque generated near stall is approximately equivalent to the energy provided by only $37\pm2$ ions per revolution, which is smaller than the previous estimate of $52$ ions, given $26$ FliG in the rotor and two ions per FliG step \cite{Sowa2005,Francis1992,Thomas2006}.

For modeling molecular motors, the Brownian ratchet models have long attracted physicists' attention since Richard Feynman popularized it a half century ago~\cite{Feynman1966,Parrondo1998,Astumian2010,Peliti2012}. Among all variants of the ratchet models (see \cite{Parrondo2002} for a review), only the isothermal chemical ratchets~\cite{Julicher1997} are relevant for biological motors. The efficiency of isothermal ratchets can reach $100\%$ under ideal conditions near equilibrium when the speed goes to zero (stall)~\cite{Parmeggiani1999}. However, the power output vanishes at this ideally efficient point, which motivates researchers to study efficiency at maximum power~\cite{Seifert2008,Esposito2009}. Another serious shortcoming of the idealized models is that realistic biological motors are under constraints on the motor-track interaction potential as well as the reversibility of the underlying chemical transitions, which can have significant effects on motoor performance. In fact, it was already realized in~\cite{Parmeggiani1999} that instead of being $100\%$ the efficiency actually {\it vanishes} at stall if spontaneous stepping transitions are included, which leaves the maximum efficiency under biological constraints an open question. 

In this paper, we address the general question on how realistic microscopic properties of the motor, such as the shape of the motor-track interaction potential, the degree of irreversibility in mechanochemical transitions, and the gating (control) of the stepping transitions affect the motor performance (efficiency, power, and maximum torque (force) generation).
We do so by developing a minimal stochastic motor model where both energy-assisted and spontaneous stepping transitions are included. The motor dynamics are determined by two intrinsic mechano-chemical functions: 1) the interaction potential of the power generating motor molecules (kinesin, myosin, or MotAB) and their counterpart track molecules (microtubule, actin, or FliG), 2) the stepping rate function that depends on the relative motor-track coordinate. Together, these two microscopic functions constitute the ``design" space of molecular motors. 
We study general thermodynamic properties of molecular engines by exploring this motor design space, where a specific motor such as BFM corresponds to one particular region. Our approach not only allows us to gain important insights on the specific molecular mechanisms for the observed properties (e.g., the torque-speed relationship for BFM). More importantly, exploring the motor design space reveals fundamental thermodynamic bounds for all realistic molecular engines and general design principles to approach these bounds.

\section{A minimal model framework for molecular motors}

The approach and terminology for the minimal motor model are based on previous modeling work on BFM \cite{Xing2006,Meacci2009}, but the general formalism can be applied to other motor systems.   
As illustrated in Fig.\,1, the interaction between stator and rotor drives the rotation of the rotor from a high potential energy position towards its equilibrium position (with the lowest potential energy). The passage of an ion enhances a stator conformational change (stepping), which brings the motor to a new stator-rotor potential where the motor is again in a high potential energy state. The newly gained potential energy continues to drive the (physical) rotation of the rotor.
This continuous process drives the system towards a sequence of new equilibrium positions and gives rise to a directed stepwise rotation \cite{Sowa2005}.

\subsection{The Fokker-Planck equation}

For a processive motor like BFM with a high duty ratio, the motor dynamics can be described by two stochastic processes: 1) the physical motion (rotation), which can be viewed as a particle sliding along an energy potential $V(\theta)$ with thermal fluctuations; 2) the chemical transitions (``stepping"), which correspond to hoping between neighboring energy potentials shifted by half a period $\theta_0$. For BFM, the stator-rotor interaction potential $V(\theta)$ has a periodicity $2\theta_0\equiv2\pi/26$, where $\theta\equiv \theta_R-\theta_S$ is the relative angle between the stator angle $\theta_S$ (``chemical" coordinate) and the rotor angle $\theta_R$ (``physical" coordinate). For a linear motor like kinesin, $V$ represents the kinesin-microtubule interaction potential with a period of $\sim 8nm$ \cite{Svoboda1993}.

 \begin{figure}
\centering
\includegraphics[width=.5\textwidth,trim={1.5cm 5.5cm 0.5cm 7.5cm},clip]{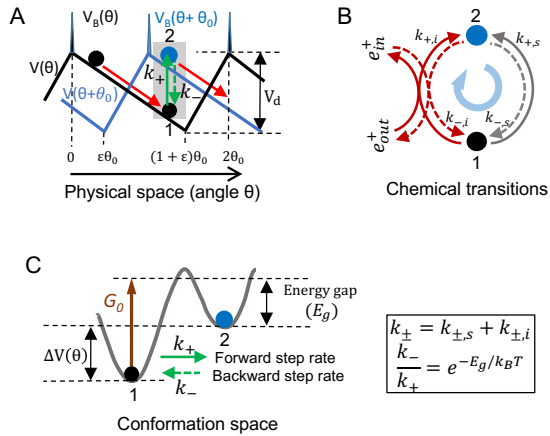}
\caption{Illustration of the minimal motor model. (A) The motor is described by its motion (red arrows) in physical space (angle $\theta$) along the interaction potential $V(\theta)$. The gray box highlights the forward and backward stepping transitions, represented by the solid and dotted green arrows respectively, between two adjacent potentials (black and blue lines) shifted by half a period $\theta_0$. A V-shaped potential is shown with its minimum at $(1+\varepsilon)\theta_0$ and depth $V_d$. An energy barrier $V_B(\theta)$ is added to prevent back flow. (B) There are two types of chemical transitions highlighted in the gray box in (A): the PMF-coupled transitions (red arrowed lines) and the spontaneous transitions (gray arrowed lines). Detailed balance is broken in the reaction loop which leads to a dissipative reaction cycle. Forward and backward reactions between state-1 to state-2 are represented by solid and dotted lines, respectively. (C) The chemical transitions shown in the chemical conformation space. The ratio between the total forward rate ($k_+$) and the total backward rate ($k_-$) depends on the energy gap $E_g$, which is the difference between the effective driving energy $G_0$ and potential gain $\Delta V(\theta)$.}
\end{figure}

Following~\cite{Xing2006}, we study the probability distribution function $P(\theta,t)$ for $\theta$ by using the Fokker-Planck (FP) equation :
\begin{equation}
\frac{\partial P}{\partial t}=\frac{\partial}{\partial\theta}[-\omega P+k_B T\xi^{-1} \frac{\partial P}{\partial\theta}]+\Delta j_s(\theta)
\label{FPE}
\end{equation}
with $\omega$ the angular speed, $\xi$ the viscous drag coefficient, and $k_BT$ the thermal energy set to $1$ hereafter. In subcellular environments, motor dynamics is over-damped and the motor speed ($\omega$) is proportional to torque: $\omega=\tau/\xi=(-V'(\theta)+\tau_{ext})/\xi$, where $\tau_{ext}$ is an external torque applied in the opposite direction of the motor rotation, $\xi$ is the viscous drag coefficient (load).

The first term on the right hand side of Eq. (\ref{FPE}) is the probability flux due to continuous physical motion. The second term $\Delta j_s(\theta)$ is the net flux due to stepping:
\begin{equation}
\Delta j_s(\theta)=\begin{cases} j_+(\theta+\theta_0)-j_-(\theta),& 0\le\theta\le \theta_0,\\  j_-(\theta-\theta_0)-j_+ (\theta),& \theta_0<\theta\le 2\theta_0, \end{cases}
\label{ss}
\end{equation}
where the forward and backward stepping fluxes are given by $j_{\pm}(\theta)=k_{\pm}(\theta)P(\theta)$ with  $k_{+}(\theta)$ the forward rate of leaving from $\theta$ to $(\theta-\theta_0)$ and $k_-(\theta-\theta_0)$ the rate of jumping back to $\theta$ from $(\theta-\theta_0)$. For simplicity, we assume $k_{+}(0<\theta<\theta_0)=0$ and $k_-(2\theta_0>\theta>\theta_0)=0$. See Sec. A and Fig.\,7 in the Appendix for details of the model derivation. 

\subsection{Irreversible chemical cycle and loose coupling}

There are two distinct pathways for chemical transitions (Fig. \,1B). For the PMF-coupled transitions, the forward transition is boosted by the PMF energy $E_0$ ($E_0$ is the ATP hydrolysis energy for linear motors), and the backward transitions regain the energy by pumping a proton out (or synthesizing ATP). The transition rates satisfy the thermodynamic constraint:
$
k_{+,i}(\theta)=e^{-\Delta V+E_0} k_{-,i}(\theta-\theta_0)
$ where $\Delta V\equiv V(\theta-\theta_0)-V(\theta)$ is the potential energy change (gain) for a forward step. There are also spontaneous transitions that are decoupled from the energy source, their rates satisfy: 
$
k_{+,s}(\theta)=e^{-\Delta V} k_{-,s}(\theta-\theta_0)
$. In the presence of both types of transitions, we have $\frac{k_{+,i}\times k_{-,s}}{k_{+,s}\times k_{-,i}}=e^{E_0}\ne 1,$ which indicates that detailed balance is broken between the chemical states (with the same physical coordinate $\theta$). Therefore,  some of the PMF energy is dissipated by the irreversible chemical reaction cycle (see Fig.\,1B) without driving any physical motion. This loss of energy prevents the system from being $100\%$ efficient.  

The relative strength of the two types of stepping transitions can be characterized by a reversibility parameter $\kappa$: 
$k_{-,i}=\kappa k_{-}, \;\; k_{-,s}=(1-\kappa)k_{-}.$ The ideal case of $\kappa=1$ corresponds to the perfectly tight-coupling scenario where every forward step transition is powered by the cheminal energy and every back step transition regains the chemical energy (pump out $H^+$ or synthesize ATP). However, most realistic molecular motors are loosely coupled (not perfectly tight-coupled) with  $0<\kappa<1$. For example, both myosin and kinesin have a net ATP hydrolysis rate at stall \cite{Bowater1988, Carter2005} and some backward steps can even cost energy \cite{Lipowsky2007}. A loose coupling mechanism is also proposed recently for BFM \cite{Boschert2015}. One of the goals of our study is to search for design principles to enhance motor performance under the realistic constraint of only partially reversible  $\kappa<1$.

Combining the two types of stepping transitions, the total transition rates $k_{\pm}(=k_{\pm,i}+k_{\pm,s})$ satisfy:
\begin{equation}
\frac{k_{+}(\theta)}{k_{-}(\theta-\theta_0)}=\exp[-\Delta V(\theta)+G_{0}]\equiv \exp(E_g),
\label{db1}
\end{equation}
where 
$G_{0}\equiv\ln(1-\kappa+\kappa e^{E_0})$ is the effective driving energy. Except for cases with extremely small $\kappa$ (we use $\kappa=0.5$ in this paper unless otherwise stated), we have $G_0 \approx E_0+\ln(\kappa)\approx  E_0$ when $-\ln\kappa\ll E_0$.  

As shown in Fig.\,1C, An energy ``gap" $E_g\equiv G_{0}-\Delta V$  is defined to characterize the difference (gap) between the effective driving energy $G_0$ and the potential energy gain $\Delta V$. From Eq. (\ref{db1}),  a positive energy gap ($E_g>0$) suppresses the back steps, which is crucial for enhancing motor efficiency as we show later in the paper. As defined, $E_g$ is $\theta$-dependent. Here, we use it to denote the energy gap at where $k_+$ is the highest. 



\subsection{Approach and general model behaviors}
Eqs. (\ref{FPE}-\ref{db1}) completely define a minimal thermodynamically consistent model for molecular motors, including linear motors like myosin, where the coordinate $\theta$ would represent the relative positional difference between myosin and actin. The steady state distribution $P_s(\theta)$ is determined by solving the steady state FP equation:
\begin{equation}
\xi^{-1}\frac{d}{d \theta}[V'(\theta)P_s(\theta)]+\xi^{-1}\frac{d^2 P_s(\theta)}{d \theta^2}+\Delta j_s(\theta)=0,
\label{ss}
\end{equation}
with periodic condition $P_s(\theta)=P_s(\theta+2\theta_0)$ and normalization $\int_0^{2\theta_0}P_s(\theta)d\theta=1$.  The intrinsic properties of the motor are characterized by two functions: the interaction potential functions $V(\theta)$ and stepping rate function $k_{+}(\theta)$ ($k_-(\theta)$ is given by Eq. (\ref{db1})). The external load is determined by $\xi$.

For a given load $\xi$, Eq. (\ref{ss}) can be solved to obtain $P_s(\theta)$, from which the average torque generated by the motor can be determined: $ \bar{\tau} (\xi) =-\int_0^{2\theta_0}V'(\theta)P_s(\theta)d\theta$, and the average (rotational) speed can be obtained by the over-damped assumption valid at low Reynolds number:  $\bar{\omega}(\xi)= \bar{\tau}(\xi)/\xi$. 
By sweeping through different values of $\xi$, the model results in a torque-speed ($\bar{\tau}-\bar{\omega}$) relationship, which can be compared directly with experiments. The maximum torque $\tau_m$ is reached at high-load ($\xi\rightarrow\infty$) when the motor is at stall ($\bar{\omega}=0$). 


In the absence of external energy source and external force, i.e., when $E_0=0$ and $\tau_{ext}=0$, the system is in equilibrium with its thermal environment. It is easy to show that the steady-state solution for Eq. (\ref{FPE}) in this case is simply the equilibrium Boltzmann distribution: $P_s(\theta)=\Omega^{-1}\exp[-V(\theta)]$, with $\Omega=\int_0^{2\theta_0}\exp[-V(\theta)]d\theta$ the normalization constant. Consequently, there is no net torque generation or motion, i.e., $\bar{\omega}=\bar{\tau}=0$. 

However, when $E_0> 0$, detailed balance is broken between different physical coordinates ($\theta$), i.e., $k_{+}(\theta)P_s(\theta)\ne k_{-}(\theta-\theta_0)P_s(\theta-\theta_0)$,  and the motor can generate a nonzero average torque to drive mechanical motion (rotation). The viscous drag $\xi\bar{\omega}$ is considered as the natural load on the motor. Even though an external torque $\tau_{ext}$ can also be applied to probe the motor behaviors, it is more convenient and biologically more realistic to change the load by varying $\xi$ as done by almost all experiments on BFM. For the remaining of this paper, we set $\tau_{ext}=0$ and varying $\xi$ except when we discuss different definitions of the motor efficiency at the end of the paper.

\section{Design principles for optimal motor performance }

In the general model framework given in the last section, the motor design space is spanned by two intrinsic functions: $V(\theta)$ and $k_+(\theta)$. For a specific motor system like BFM, specific choices of $V(\theta)$ and $k_+(\theta)$ were made to fit experimental data and backward transitions $k_-(\theta)$ were typically neglected. 
Here, we treat $k_+(\theta)$ and $V(\theta)$ as a variable functional, and we always keep $k_-(\theta)$, which is determined from $k_+(\theta)$ and $V(\theta)$ by using Eq. (\ref{db1}). 
By systematically exploring the motor design space, our main goal is to investigate fundamental limits and possible design principles for optimal motor performance characterized by its power output and energy efficiency for a given driving energy $E_0$.

\subsection{A gating mechanism for high power generation}


The average power output of the motor, defined as $\dot{W}=\bar{\tau}\bar{\omega}$, can only be high if both $\bar{\tau}$ and $\bar{\omega}$ are high. The measured torque-speed curve for CCW BFM has a concave down shape with a roughly constant high torque at low to medium speeds and a fast decrease of torque at high speeds \cite{Chen2000,Lo2013}. This concave torque-speed curve has the advantage of generating high power output (or equivalently a high torque for a given speed) in a wide range of physiologically relevant loads. Here, we study the general design requirements for such a concave torque-speed dependence, which is critical for high power generation.

The form of the periodic potential $V(\theta)$ is characterized by two parameters: the depth of the potential $V_d$, and the location of its minimum $\theta_m\equiv(1+\varepsilon)\theta_0$, where $\varepsilon\in[0,1]$ is an asymmetry parameter. A symmetric potential corresponds to $\varepsilon=0$, and $\varepsilon=1$ represents the extreme case when the potential is infinitely steep at $\theta=2\theta_0$.  For simplicity, we used a  piece-wise linear form ($V$-shaped) for $V(\theta)$ for most of the paper as shown in Fig.\,1A and Fig.\,2A. Other forms of $V(\theta)$, such as parabolic functions, were also used without affecting the main results (see Section C and Fig.\,9 in Appendix for details). For the $V$-shaped potential, the torque generated from this potential is positive $\tau(\theta)=\tau_+ \equiv V_d/\theta_m>0$ for $0\leq \theta<\theta_m$, and negative $ \tau(\theta)=-\tau_- \equiv -V_d/(2\theta_0-\theta_m)<0$ for $\theta_m<\theta \leq2\theta_0$, as shown in Fig.\,2A.
A high energy barrier $V_B$ near the peak of $V(\theta)$ is also added to prevent slipping between two adjacent FliG's without stepping. A piece-wise linear form of $V_B$ is used (see Appendix B).  
In the following, we focus on elucidating the role of controlling (gating) the stepping transitions, i.e., the specific form of $k_+(\theta)$, for obtaining the observed torque-speed characteristics and high power generation. 

\subsubsection{ An analytical solution for the torque-speed relationship} 

We derive an approximate analytical solution for the torque-speed curve from our model based on ideas introduced before \cite{Meacci2009, Mandadapu2015}. At a microscopic timescale, the motor moves in two alternating modes: moving and waiting. The moving phase corresponds to the duration when the motor moves down the potential $V$ and generates a positive torque $\tau_+(>0)$. The average moving time is approximately $\langle t_m \rangle \approx \xi \theta_0/\tau_+$. The waiting phase begins when the motor reaches the potential minimum $\theta_m$. The waiting phase may be skipped due to stepping and the probability of reaching the potential minimum is $p_w=\exp(-\xi K/\tau_+)$, where $K\equiv \int_{\theta_0}^{\theta_m}k_+(\theta)d\theta$ is the integrated forward stepping rate over $[\theta_0,\theta_m]$. Once reaching $\theta_m$, the motor fluctuates (due to thermal noise) around $\theta_m$ ``waiting" for the next stepping transition to occur. During the waiting phase, the system approximately follows the equilibrium distribution $P_s(\theta)\approx \Omega^{-1}\exp[-V(\theta)]$. So the average waiting time $\langle t_w\rangle \approx k_0^{-1}$, where $k_0\approx \Omega^{-1}\int_0^{2\theta_0}k_+(\theta)\exp[-V(\theta)]d\theta$ is the average stepping rate in the waiting phase. By combining these considerations, we obtain an approximate solution for the speed $\bar{\omega}\approx \theta_0(\langle t_m\rangle+p_w \langle t_w\rangle)^{-1}$.
By introducing a re-scaled torque $\tilde{\tau}\equiv \bar{\tau}/\tau_+$ and a re-scaled speed  $\tilde{\omega}\equiv \bar{\omega}/\omega_m$ with $\omega_m(=k_0\theta_0)$ the maximum speed, we obtain an approximate analytical expression for the torque speed curve:
\begin{equation}
\tilde{\tau}+\tilde{\omega} \exp(-\frac{q\tilde{\tau}}{\tilde{\omega}})=1,
\end{equation}
with a single parameter $q$ that depends on $V(\theta)$ and $k_+(\theta)$:
\begin{equation}
q\equiv \frac{K}{k_0}=\frac{(\int_{\theta_0}^{\theta_m} k_+(\theta)d\theta)\times ( \int_0^{2\theta_0}\exp[-V(\theta)]d\theta)}{\int_0^{2\theta_0}k_+(\theta)\exp[-V(\theta)]d\theta}.
\label{q_exp}
\end{equation}
The concavity of the torque-speed curve is determined by $q$. For $q\rightarrow 0$, torque-speed curve is linear $\tilde{\tau}+\tilde{\omega}=1$ with zero concavity. As $q$ increases, the concavity increases.

What is the design of $k_+(\theta)$ that gives rise to a large value of $q$ for a given $V(\theta)$? The answer is revealed by Eq. (\ref{q_exp}). For the $V$-shaped potential, the dependence of $K$ and $k_0$ on $k_+(\theta)$ shows that higher stepping rates in a narrow region away from the potential minimum can increase $K$ without increasing $k_0$ too much and thus lead to a larger value of $q$.   
This ``gating" region characterized by a small width $\theta_g(\ll\theta_0)$ and a large stepping rate $k_g(\gg k_0)$ within the interval $(\theta_0,\theta_m)$ but closer to $\theta_0$, serves to prevent the motor from entering the waiting phase at high loads without increasing the maximum speed at low loads. These effects of the gating mechanism lead to the observed concavity in the torque-speed curve.


\subsubsection{Simulation results} 

We verified this gating mechanism by direct numerical simulations. For simplicity, we choose a piecewise constant profile for $k_+$ as shown in Fig.\,2A: 1) $k_+(\theta)=k_g$ for $\theta\in[\theta_0+\theta_{\varepsilon}, \theta_0+\theta_{\varepsilon}+\theta_g)$; 2) $k_+(\theta)=k_a$ for $\theta\in[\theta_0+\theta_{\varepsilon}+\theta_g,\theta_m)$; 3) $k_+(\theta)=k_b$ for $\theta\in[\theta_m, 2\theta_0-\theta_{\varepsilon})$; and zero otherwise. 
Here, $\theta_{\varepsilon}(>0)$ controls the gate location, $\theta_g$ and $k_g$ are the width and stepping rate of the gate region, $k_a$ and $k_b$ represent the background stepping rates to the left and right of the potential minimum, respectively.

For a given $k_+(\theta)$, we solve Eq. (\ref{ss}) numerically to determine the steady state distribution $P_s(\theta)$ for any given load $\xi$. 
As shown in Fig.\,2B, at high ($\xi=1$, red line), $P_s(\theta)$ is mainly concentrated in the positive-torque region due to the gating effect, while it shifts to mostly populate  around the potential bottom ($\theta_m$) at low load ($\xi=0.01$, green line), and it behaves somewhere in between for intermediate load ($\xi=0.1$, blue line).  
We have computed the torque-speed curve for different values of $k_g$. As shown in Fig.\,2C, the concavity disappears as $k_g$ decreases. Note that for flagellar motor, we usually plot torque versus speed instead of speed versus external applied force as typically done in the linear motor case.   
The positioning of the gate is also studied. The concavity increases as the gate is moved away from the potential minimum at $\theta_m$ towards the midpoint at $\theta=\theta_0$, i.e., as $\Delta _g \equiv \theta_m-(\theta_0+\theta_{\varepsilon})$ increases, as shown in Fig.\,2D. The dependence of the concavity of the torque-speed curve on the strength and position of the gate, as shown in Fig.\,2C\&D, agrees with our analytical results.

The normalized torque-speed curve with a strong gating strength and proper positioning (the red lines in Fig.\,2C\&D) agrees with experimental data \cite{Lo2013} for the CCW BFM (square symbols in Fig.\,2C\&D). The predicted dependence of concavity on the gating mechanism also provides a possible mechanism for the CW motor, which shows a linear torque-speed curve \cite{Yuan2010}. These predicted dependence may be tested by future experiments that measure the torque-speed curve in cells with mutated residues around their ion channel \cite{Blair1991,Blair1999}.  
\begin{figure}
\centering
\includegraphics[width=.45\textwidth]{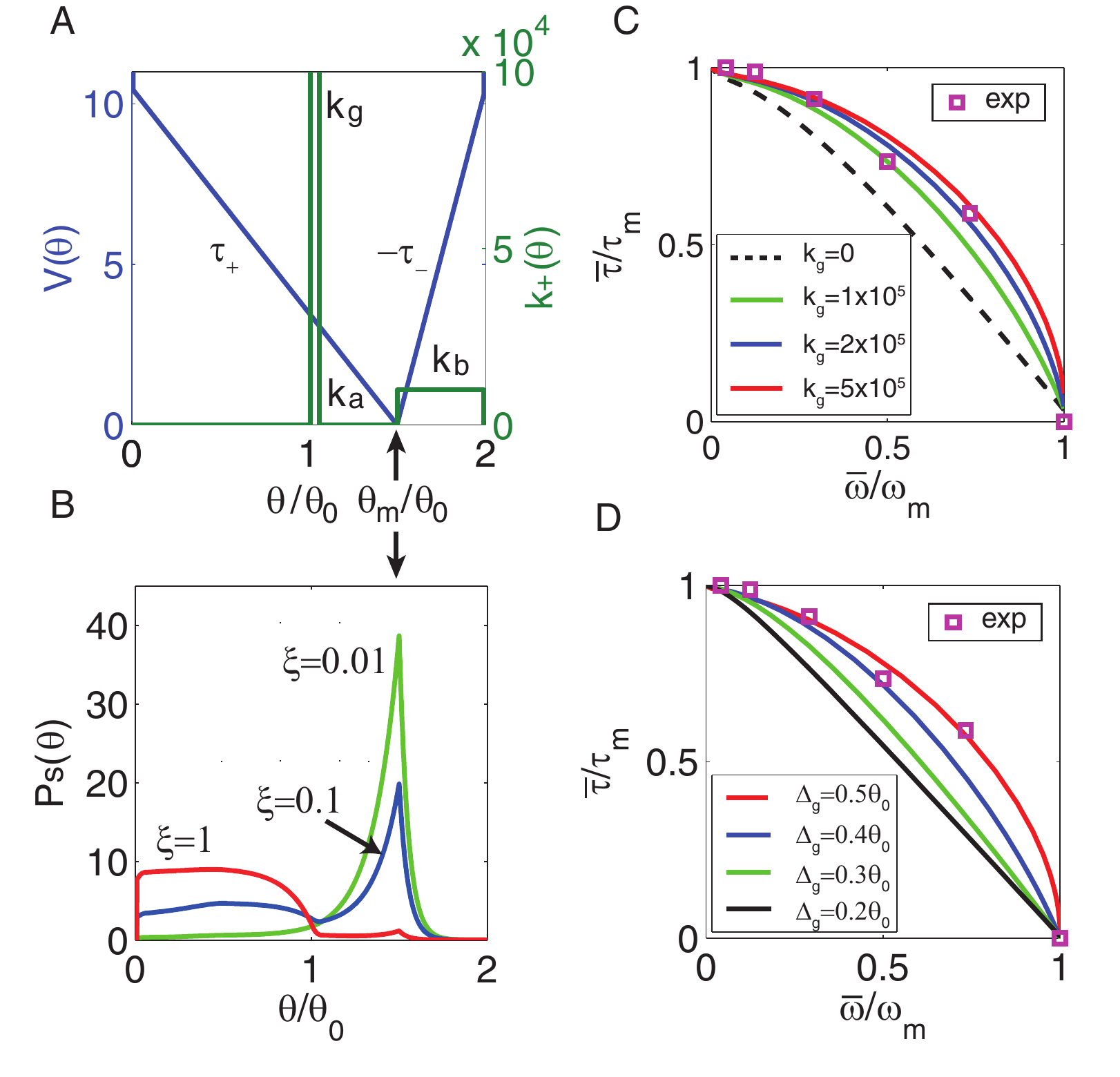}
\caption{The motor probability distribution and the gating effect on the torque-speed curve. (A) The stator-rotor interaction potential $V(\theta)$ (blue line); and the forward stepping rate $k_+(\theta)$ (green line). A positive torque $\tau_+$ is generated when $\theta<\theta_m$ and a negative torque $-\tau_-$ is generated when $\theta>\theta_m$. (B) The steady-state distribution $P_s(\theta)$ at three representative loads: high load $\xi = 1$ (red), medium load $\xi=0.1$ (blue), and low load $\xi=0.01$ (green) for $k_g=5\times 10^5$ with $V(\theta)$ and $k_+(\theta)$ given in (A). Note that the peak of $P_s(\theta)$ at the bottom of potential $\theta_m$, indicated by the arrows in both (A) and (B), increases as the load ($\xi$) decreases. 
(C) The torque-speed curves for different values of the gating strength $k_g$. The concavity increases with the gating strength $k_g$.  (D) The torque-speed curves for different values of the distance $\Delta_g\equiv \theta_m-(\theta_0+\theta_{\epsilon})$ between the gate and the potential minimum. The concavity increases with $\Delta_g$. The square symbols in both (C)\&(D) represent data from \cite{Lo2013} (pH=7.0, $[Na]_{ex}=30mM$). }
\end{figure}

\subsection{The maximum torque at stall is limited by speed fluctuations}

Another important characteristic of any molecular motor is the maximum torque $\tau_{max}$ (or maximum force for a linear motor) that the motor generates near stall. For a given $G_0$, we ask the question what is the best design of $V(\theta)$ that optimizes $\tau_{max}$. Naively, it may be desirable to have a steep interaction potential to generate a large $\tau_{max}$. In the case of the $V$-shaped potentials, one would expect $\tau_{max}$ to increase with the gradient ($\tau_+$) of the potential. We have computed $\tau_{max}$ in our model for different choices of $\tau_+$. Surprisingly, as shown in Fig.\,3A, although $\tau_{max}$ increases with $\tau_+$ for small $\tau_+$, it reaches a peak value $\tau_{max}^p<G_0/\theta_0$ at a finite $\tau_+=\tau_+^p<G_0/\theta_0$ and decreases sharply for $\tau_+>\tau_+^p$.

What causes this non-monotonic dependence of $\tau_{max}$ on $\tau_+$? For a larger value of $\tau_+$, the torque generated in the positive torque regime ($\theta<\theta_m$) is larger. However, the backward stepping rate is also higher as the energy gap $E_g=G_0-\tau_+\theta_0$ is lower. The higher backward stepping rate increases the probability in the negative torque regime ($\theta>\theta_m$) and thus decreases the average torque (see Appendix C and Fig.\,8 for details). These two competing effects of varying $\tau_+$ lead to the existence of a maximum $\tau_{max}$. Different choices of $\varepsilon$ only change the peak slightly without changing the general behavior of $\tau_m$ (Fig.\,3A). 

\begin{figure}
\centering
\includegraphics[width=.45\textwidth]{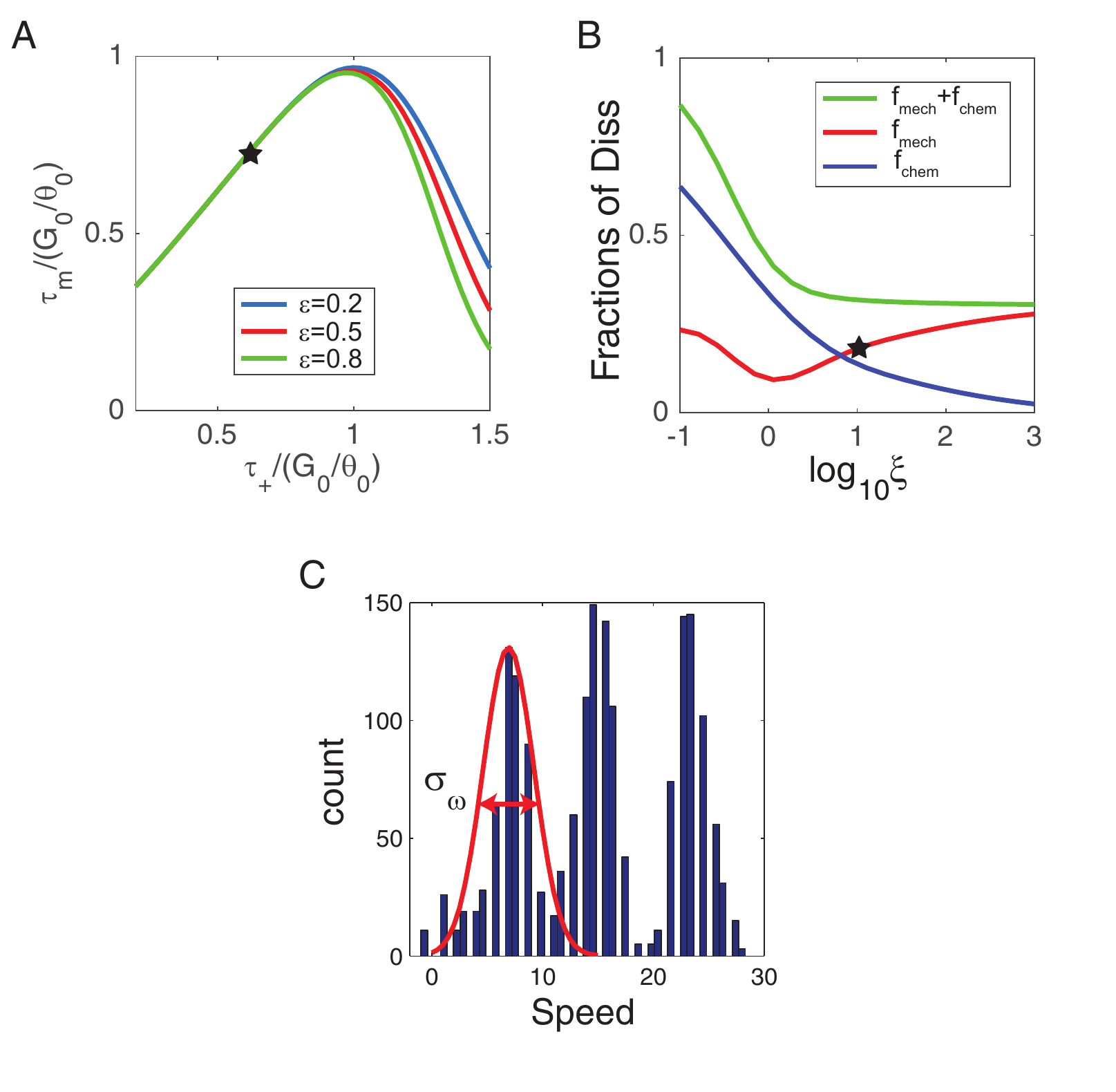}
\caption{Maximum torque $\tau_m$ and the energy dissipations. (A) $\tau_m$ depends non-monotonically on the potential gradient $\tau_+$ for different values of $\varepsilon=0.2,0.5,0.8$. The experimentally measured $\tau_m$ and its corresponding $\tau_+$ is marked by the star. (B) Fractions of energy dissipation due to torque (speed) fluctuations $f_{mech}$ (red line) and entropy production in chemical reactions $f_{chem}$ (blue line) versus load ($\xi$).  $f_{mech}$ dominates at high loads, $f_{chem}$ dominates at low loads. The total dissipation $(f_{mech}+f_{chem})$ is shown as the green line. (C) The experimentally measured speed distribution when hook-only motors were attached to a large 1$\mu m$ polystyrene bead with an estimated high load of $\xi\approx 8$ (see \cite{Lo2013} for details). The variance around the first peak $\sigma_{\omega}$ corresponds to the speed fluctuation for motors with a single stator. The fractional dissipation due to speed fluctuation at $\xi=8$ can be estimated $f_{mech}\approx 0.16$ (marked as a star) from (C).  The original data are kindly provided by Dr. C-J Lo \cite{Lo2013}.}
\end{figure}

\subsubsection{Thermodynamic laws for molecular motors} 

The bound for $\tau_{max}$ can be obtained rigorously by studying the thermodynamic torque $\tau_l(\theta)=-(V(\theta)+\ln P_s(\theta))'$, where the first term represents the torque from the stator-rotor interaction and the second term is the ``entropic" torque from thermal fluctuations akin to the thermodynamic pressure. 
By integrating the steady state Fokker-Planck equation, we obtain the average $\tau_l$:
\begin{equation}
\langle \tau_l\rangle=\xi \theta_0\int_{0}^{2\theta_0}[j_+(\theta)-j_-(\theta)]d\theta
=\xi \theta_0(J_+-J_-),
\label{w_Dn}
\end{equation}
where $J_{\pm}\equiv \int_{0}^{2\theta_0}j_{\pm}(\theta) d\theta$ are the total forward and backward fluxes. 
The second moment of $\tau_l$ can be computed:
$\langle\tau_l^2\rangle=\int_0^{2\theta_0}\tau_l^2(\theta)P_s(\theta)d\theta=-\int_0^{2\theta_0}(V+\ln P_s)(V'P_s+P_s')'d\theta,$
where boundary terms are set to zero. In steady state, Eq. (\ref{FPE}) leads to: $(V'P_s+P_s')'=-\xi \Delta j_{s}$. By using Eq. (\ref{ss}) for $\Delta j_s$ and Eq. (\ref{w_Dn}) for $\bar{\tau}$, we have:
\begin{equation}
\langle\tau_l^2\rangle=\frac{G_0\bar{\tau}}{\theta_0}-\xi S_j,
\label{Stau}
\end{equation}
where $S_j\equiv \int_0^{\theta_0}[j_+(\theta+\theta_0)-j_-(\theta)]\ln\frac{j_+(\theta+\theta_0)}{j_-(\theta)}d\theta $ is the entropy production rate of the chemical reactions.

In steady state, the power output or the rate of mechanical work performed by the motor (against viscous drag) is $\dot{W}\equiv \bar{\omega}\bar{\tau}$. Using Eq. (\ref{Stau}), we derive an equation for $\dot{W}$:
\begin{equation}
\frac{G_0\bar{\omega}}{\theta_0}=\bar{\omega}\bar{\tau}+\xi^{-1}\sigma_{\tau}+S_j,
\label{FirstLaw}
\end{equation}
where $\sigma_{\tau}\equiv \langle\tau_l^2\rangle -\bar{\tau}^2$ is the variance of the thermodynamic torque. 

Eq. (\ref{FirstLaw}) is the first law of thermodynamics for a nonequilibrium motor system with an external energy source. The left hand side of Eq. (\ref{FirstLaw}) represents the rate of energy input. The first term on the right hand side (RHS) of Eq. (\ref{FirstLaw}) represent the average power output. In addition, there are two distinct sources of energy dissipation. $S_j$ is the energy loss due to entropy production and the corresponding heat generation in {\it chemical space}. $\xi^{-1}\sigma_{\tau}$ is the energy dissipation due to fluctuations of torque and speed in {\it physical space}. We note that the speed and torque fluctuations depend on the non-equilibrium motor dynamics (driven by $G_0$) in addition to thermal noise. In particular, the torque fluctuation $\sigma_{\tau}$ is finite even when temperature goes to zero.  

The second law of thermodynamics for the motor manifests itself by the fact that these two energy dissipation rates are positive definite:
\begin{equation}
S_j \ge 0, \;\;\;\xi^{-1}\sigma_{\tau}\ge 0.
\label{SecondLaw}
\end{equation}
From the first and second law, Eqs. (\ref{FirstLaw}-\ref{SecondLaw}), it follows that the average torque is bounded:
\begin{equation}
\bar{\tau}=G_0/\theta_0-\sigma_{\tau}/\bar{\tau}-\xi S_j/\bar{\tau}\le G_0/\theta_0.
\end{equation}

\subsubsection{Simulation results and experiments} 

The question now is whether the maximum torque $\tau_{max}$ can ever reach this theoretical limit $G_0/\theta_0$. At high load $\xi\gg 1$, the entropy production rate is small $S_j\propto \xi^{-2}$ because both $\ln[\frac{j_+(\theta+\theta_0)}{j_-(\theta)}]\propto\xi^{-1}$ and $ [j_+(\theta+\theta_0)-j_-(\theta)]\propto\xi^{-1}$. However, in general $\sigma_{\tau}$ does not vanish in the high load limit. The torque variance $\sigma_{\tau}$ depends on the shape of $V(\theta)$ and only approaches zero when the interaction potential takes the extreme limit of $\varepsilon \rightarrow 1$ with delta-function energy barrier. Given the size of a motor protein ($\sim 4 nm$) and that of a typical amino acid ($\sim 0.8nm$), the asymmetry parameter should be $\varepsilon< 1-0.8/(2\times 4)=0.9$. Therefore, any realistic form of $V(\theta)$ results to a finite $\sigma_{\tau}$ and thus a maximum torque that is less than $G_0/\theta_0$.

We have computed $\bar{\tau}$, $\bar{\omega}$, $\sigma_{\tau}$, and $S_j$ for different load ($\xi$) in our model numerically. The fraction of energy dissipation through speed fluctuation and entropy production are given by $f_{mech}\equiv \sigma_{\tau}/(\xi\bar{\omega}G_0/\theta_0)$ and $f_{chem}\equiv S_j/(\bar{\omega}G_0/\theta_0)$, which are shown in Fig.\,3B as red and blue lines respectively. Consistent with our analysis, the dissipation due to speed fluctuation $f_{mech}$ reaches a nonzero constant as $\xi\rightarrow\infty$ while the dissipation from entropy production $f_{chem}\rightarrow 0$. 

In the recent experiments by Lo et al. \cite{Lo2013}, the maximum torque near stall was found to be $\tau_{max}\approx 71\% \frac{G_0}{\theta_0}$. From our analysis, this means that at least $29\%$ of IMF is dissipated, and an increasing portion of the dissipation is caused by speed and torque fluctuations as the load increases (see red line in Fig.\,3B). In Fig.\,3C, the experimentally observed speed distribution at a high load ($1\mu m$ bead) \cite{Lo2013} is shown. Consistent with our analysis, significant speed fluctuations are present. Quantitatively, the average and variance of motor speeds for the motors with a single stator (those speeds around the first peak in Fig.\,3C) are estimated to be $\bar{\omega}\approx 6.5 Hz$ and $\sigma_{\omega}\approx 6.4 Hz^2$.  The fraction of energy dissipation due to speed fluctuation can be estimated: $\sigma_{\tau}\theta_0/(\bar{\tau}G_0)\approx \sigma_{\omega}/\bar{\omega}^2\times \tau_{max}\theta_0/G_0\approx 6.4/6.5^2\times 0.71 = 0.11$, which is in the same range but lower than the value $0.16$ obtained from our model at the corresponding load (marked by a star in Fig.\,3B). The reason for this quantitative difference may be that the model result depends on the detailed shape of $V(\theta)$, which is not tuned in this study. Additionally, $\sigma_{\omega}$ may be an underestimate of the instantaneous speed fluctuation due to the experimental averaging process. Future experiments with high temporal resolution are needed to measure dynamics of the instantaneous speed fluctuation and to compare it directly with our model prediction in order to understand the microscopic origin of speed fluctuation and energy dissipation.


\subsection{Performance limits in loosely coupled motors ($\kappa<1$) } 


The motor's power output is given by $\dot{W}=\bar{\omega}\bar{\tau}=\theta_0 (J_+-J_-)\bar{\tau}$. To determine the motor efficiency, we need to know the net free energy cost. Since only the proton-assisted transitions $k_{\pm,i}$ are coupled with energy consumption and regeneration, the average net energy consumption rate is:
$\Delta G(\kappa) = E_0(J_{+,i}-J_{-,i})\approx E_0(J_{+}-\kappa J_{-}),$
where we have neglected the much smaller spontaneous forward flux $J_{+,s}=(1-\kappa)e^{-G_{0}}J_{+}\ll J_{+}$. 
The motor efficiency can then be defined accordingly:
\begin{equation}
\Lambda(\kappa)=\frac{\dot{W}}{\Delta G (\kappa)}=\frac{\bar{\tau}\theta_0}{E_0} \frac{J_+-J_-}{J_+-\kappa J_-}.
\label{eff1}
\end{equation}


\subsubsection{Maximum efficiency occurs at a finite speed with a positive energy gap} 

We have computed both the power output ($\dot{W}$) and efficiency ($\Lambda(\kappa)$) in our model for different choices of interaction potential $V(\theta)$ characterized by $E_g$ (equivalently $\tau_+$ or $V_d$). 
As expected, $\dot{W}=\bar{\tau}\times\bar{\omega}$ reaches its maximum value $\dot{W}_m$ at a finite load (or a finite speed) and a positive energy gap $E_g>0$. Surprisingly, however, for a loosely coupled motor with $\kappa<1$, the efficiency $\Lambda$  shows a similar behavior with its maximum at a finite load (or finite speed) as shown in Fig.\,4B. 

The efficiency-speed dependence is further studied for different values of $\kappa$. As shown in Fig.\,4C, for high speeds (or low loads) $\Lambda$ is independent of $\kappa$ and decreases with speed. A strong dependence on $\kappa$ occurs at low speeds (high loads). For {\it any} value of $\kappa < 1$, instead of reaching its maximum at zero speed, the efficiency vanishes linearly with speed. Only in the singular case of $\kappa=1$, does $\Lambda$ reach its maximum value at zero speed. In any loose-coupling motors ($\kappa<1$), the efficiency $\Lambda$ reaches its maximum at a finite speed. This is a much more ``useful" maximum efficiency as the power output can also be high unlike the case of the purely reversible motor with $\kappa=1$ where the maximum efficiency occurs at zero power.

\begin{figure}
\centering
\includegraphics[width=.45\textwidth]{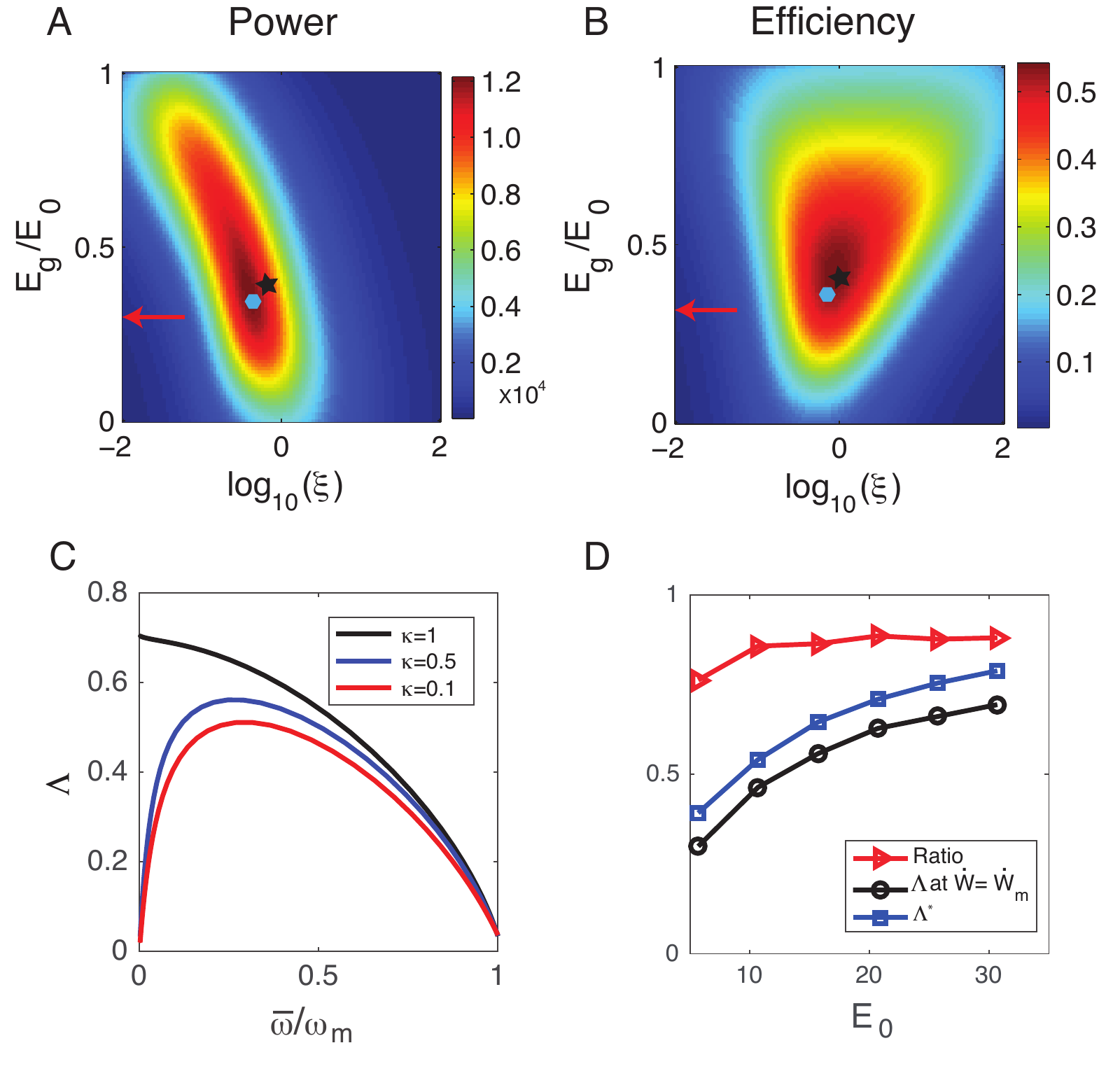}
\caption{Power and efficiency of the motor. The dependence of (A) power, and (B) efficiency (for $\kappa=0.5$) on energy gap $E_g$ and load $\xi$. Both power and efficiency peak at an intermediate load and $E_g$ , labeled by the blue dot (for power) and the black star (for efficiency). The red arrow indicates the $E_g/E_0\approx 0.29$ estimated from experiments \cite{Lo2013}, which is close to the optimal $E_g/E_0$ ratios for maximum power (blue dot) and maximum efficiency (black star). (C) Efficiency $\Lambda$ versus normalized speed for different values of $\kappa$. $\Lambda$ vanishes at $\bar{\omega}=0$ for all values of $\kappa<1$. (D) The efficiency of the motor working at maximum power (blue line) is comparable to the (global) maximum efficiency $\Lambda^*$ (black line), with their ratio (red line) $\sim 80\%$ for a wide range of $E_0$. }
\end{figure}
  
To determine whether the motor can operate in a regime with both high efficiency and high power, we computed the efficiency at the maximum power, $\Lambda(\dot{W}=\dot{W}_m)$, and the global maximum efficiency $\Lambda^*$ in our model for different $E_0$ (Note that we explore the whole range of load and power output instead of just focusing on the efficiency at the maximum power \cite{Peliti2012}).  As shown in Fig.\,4D, the ratio, $\Lambda(\dot{W}_m)/\Lambda^*$, is as high as about $80\%$ for a wide range of $E_0$. This means that the rotary motor can {\it simultaneously} achieve both high efficiency and high power output, which is evident from the closeness of the peak positions for $\dot{W}$ and $\Lambda$ shown in Fig.\,4A\&B. Indeed, the value of $E_g/E_0\approx 0.29$ estimated from experimental data \cite{Lo2013}, marked by the red arrowed line in Fig.\,4A\&B, is close to the optimal $E_g/E_0$ ratios for maximum power (blue dot) and maximum efficiency (black star). 

Both power and efficiency depend non-monotonically on the energy gap $E_g$, as shown in Fig.\,4A\&B. On one hand, a large energy gap can suppress backward steps since $k_-(\theta)=k_+(\theta+\theta_0)e^{-E_g}$. On the other hand, since the system gains a potential energy $\Delta V\equiv V(\theta)-V(\theta-\theta_0)=G_0-E_g$, which converts to mechanical work during the subsequent power stroke, a larger $E_g$ means a smaller work performed by the forward steps. This tradeoff leads to the non-monotonic dependence on $E_g$ and an optimal motor performance (power and efficiency) at a positive finite $E_g$. 

We have determined the maximum efficiency $\Lambda^*$ at different $E_0$ for different $\kappa$ and $\varepsilon$ numerically. 
Remarkably, as shown in Fig.\,5A, the maximum efficiency $\Lambda^*$, though less than $1$, can reach a high value even when most of the back steps are spontaneous, i.e., when $\kappa$ is small (e.g., $0.1$). In fact, $\Lambda^*$ can approach $1$ as $E_0\rightarrow\infty$ and the difference $(1-\Lambda^*)$ is found to scale with $E_0$ as $\ln(E_0)/E_0$ (to the leading order) for $E_0\gg 1$ : 
\begin{equation}
1-\Lambda^*=C_e(\kappa,\varepsilon)\times \frac{\ln (E_0)}{E_0}+h.o.t.,
\label{scale}
\end{equation}
where $C_e$ is a prefactor that only depends on $\kappa$ and $\varepsilon$. Note that energy is expressed in unit of $k_BT$, and $E_0$ should be understood as $E_0/k_B T$ in the above expression.

Intuitively, the optimal efficiency $\Lambda^*$ is reached by balancing two opposing effects of $E_g$ as mentioned before. 
A naive design of $V(\theta)$ would be to have a large positive torque $\tau_+$ given by the driving energy and the step size, $\tau_+=G_0/\theta_0$. However, this naive design would lead to $E_g=0$ and thus a high value of $k_-$, which lowers the motor efficiency when $\kappa<1$.  Given that $k_-/k_+=e^{-E_g}$ depends exponentially on $E_g$ (Eq. (\ref{db1})), the maximum efficiency shown in Fig.\,5A is achieved with the choice of a small but positive energy gap $E_g^*$ that depends (roughly) logarithmically on $E_0$  as shown in Fig.\,5B, which is the origin of the logarithmic dependence in Eq. (\ref{scale}). The prefactor $C_e$ in Eq. (\ref{scale}) is an order $1$ constant and decreases weakly with $\kappa$ for $\kappa\le 0.95$ as shown in Fig.\,5C. It decreases sharply only near $\kappa=1$, but remains finite even at $\kappa=1$ due to the limit on $\tau_{m}$ discussed before. $C_e$ goes to zero only at the doubly unrealistic case of having both $\varepsilon=1$ and $\kappa=1$. 

\begin{figure}
\centering
\includegraphics[width=.5\textwidth]{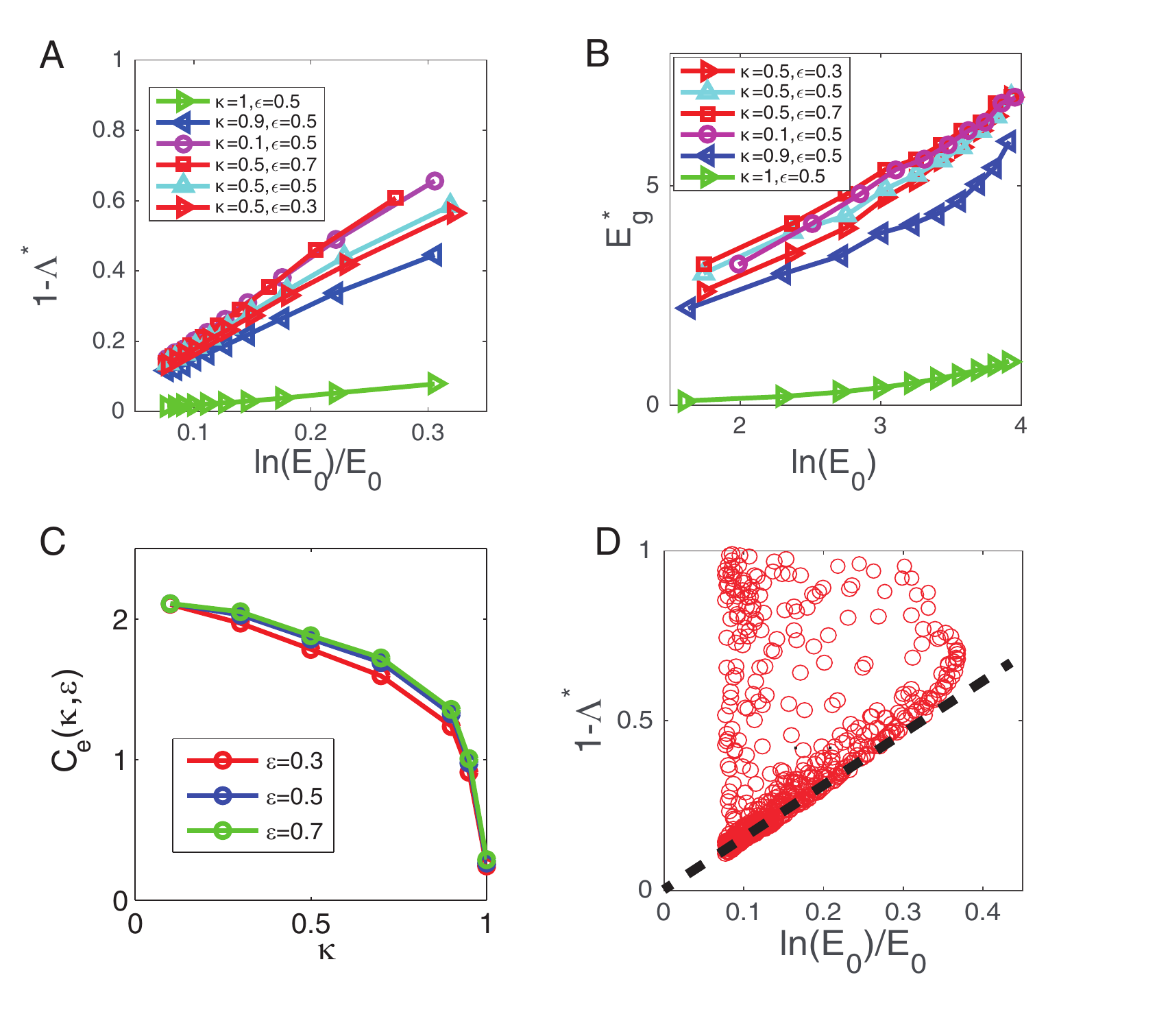}
\caption{Maximum efficiency and optimal design of the motor. (A) The scaling relationship, $1-\Lambda^*\sim \ln(E_0)/E_0$, holds for different values of $\varepsilon$ and $\kappa<1$. (B) The optimal energy gap $E*_g$ for achieving the maximum efficiency shown in (A). (C) The prefactor $C_e$ in the scaling relation, Eq. (\ref{scale}), decreases with increasing $\kappa$ for different values of $\varepsilon$. (D) $1-\Lambda^*$ versus $\ln(E_0)/E_0$ for randomly chosen motor designs. Each point corresponds to a random stepping rate profile (see Appendix B for details) with $\varepsilon=0.5$ and $\kappa=0.5$. All points lie above the envelop line of $1-\Lambda^*\sim\ln(E_0)/E_0$.}
\end{figure}

To verify the robustness of the maximum efficiency result (Eq.(\ref{scale})), we performed an extensive search in the motor design space. In particular, we randomly selected the three parameters $\{k_g, k_a, k_b\}$ for $k_+(\theta)$ with $\log_{10}k_g\in[0, 4],\log_{10}k_a\in[0, 3],$ and $\log_{10}k_b\in[0, 2]$ uniformly sampled. For a given $k_+(\theta)$ profile, we determined the maximum efficiency for different choices of $V(\theta)$ by varying $E_g$. In Fig.\,5D, each point represents the maximum efficiency for a random $E_0\in[1,50]k_BT$ for a random $k_+(\theta)$ function.  
As evident from Fig.\,5D, a limiting envelope (the dotted line) emerges with the highest efficiency $\Lambda^*$ following the same dependence on $E_0$ as given in Eq. (\ref{scale}): $1-\Lambda^*\propto \ln(E_0)/E_0$ for large $E_0\gg 1$.

\subsubsection{Efficiency in the presence of external forcing} 

For most of our study here, we set the external applied torque (force) $\tau_{ext}=0$ and change the load by varying $\xi$. The power of the motor, $\dot{W}=\bar{\tau}\times\bar{\omega}$, is used to overcome the viscous drag force of the load and the efficiency defined by using this power defintion is called the Stokes efficiency by Wang and Oster \cite{Wang2002}. For $\tau_{ext}\ne 0$, the output power delivered to overcome this fixed extenal torque is $\dot{W}_e\equiv \tau_{ext}\bar{\omega}$, the effciency based on $\dot{W}_e$ is the so called ``thermodynamic" efficiency \cite{Parmeggiani1999,Zimmermann2012}. Both the thermodynamic efficiency and the Stokes efficiency are well defined in the sense that they are both less or equal than $1$. However, in most biological systems there is no active component exerting a fixed force (or torque) on the molecular motor. Instead, a motor needs to overcome a passive drag force from the attached cargoes (loads) in the highly viscose cellular environment with low Reynolds number. Nonetheless, our model can be used to study the thermodynamic efficiency $\Lambda_T(\kappa)\equiv \dot{W}_e/D(\kappa)$ with $\dot{W}_e\equiv \tau_{ext}\bar{\omega}$ by varying $\tau_{ext}$ while fixing $\xi$ to be a small value (we take $\xi=0.1$ here). As shown in Figure 6, the peak efficiency  occurs at an intermediate $\tau_{ext}$ and with a finite gap $E_g$ in the potential to prevent wasteful back steps (Fig.\,6A). The dependence of the maximum thermodynamic efficiency $\Lambda^*_T$ on the driving energy $E_0$ (Fig.\,6B) also follows the same general trend as for the Stokes efficiency (Fig. \,4D and Fig.\,5C). 

\begin{figure}
\centering
\includegraphics[width=.5\textwidth]{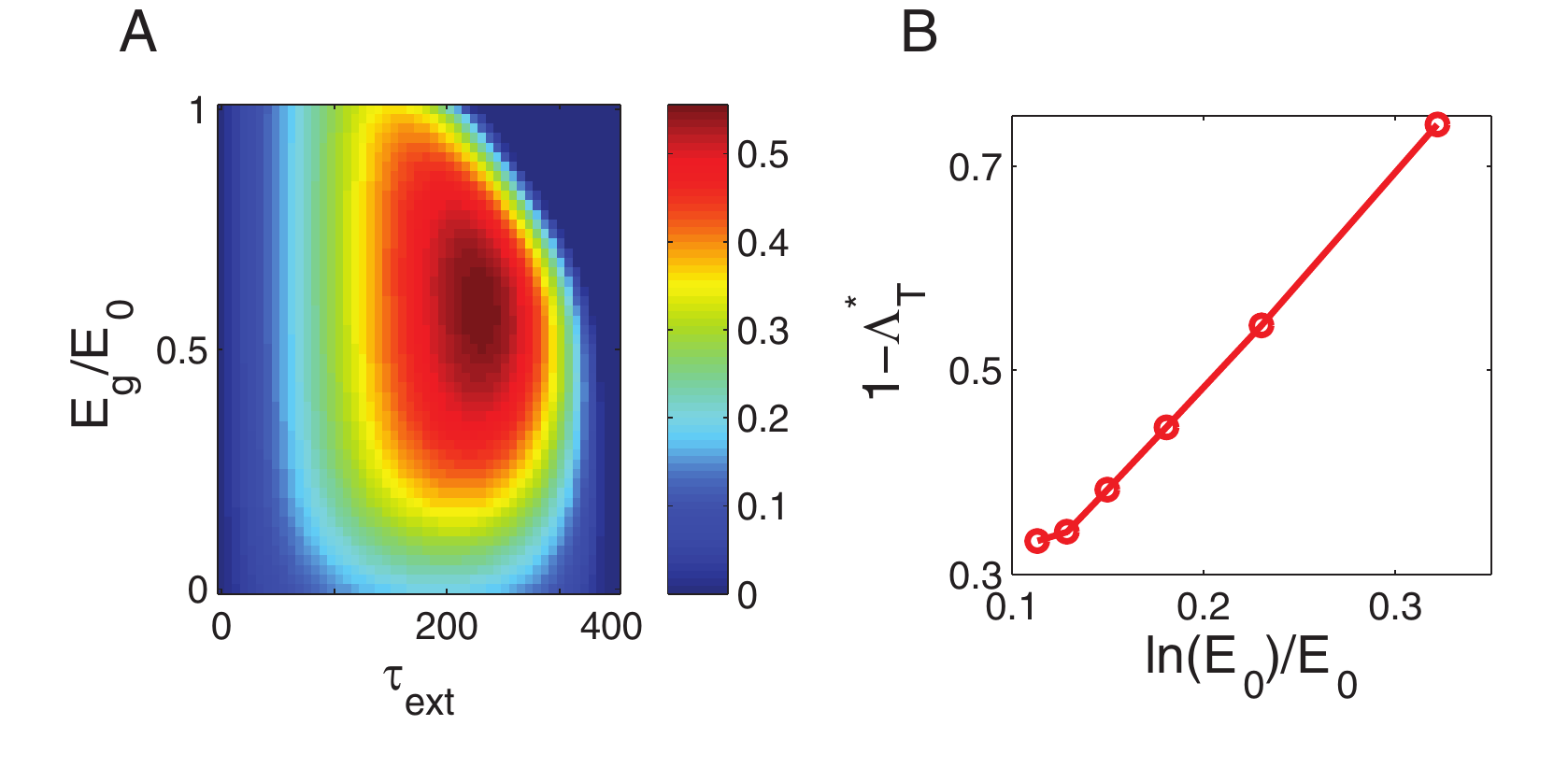}
\caption{The thermodynamic efficiency $\Lambda _T$ and its limit. (A) The dependence of $\Lambda_T$ on the energy gap $E_g$ and the external torque $\tau_{ext}$. The load $\xi=0.1$ is fixed. Other parameters are the same as those in Fig.\,4. The optimal $\Lambda_T$ occurs at a finite $E_g$ and an intermediate $\tau_{ext}$. (B) The optimal $\Lambda_T^*$ depends on the driving energy $E_0$ approximately following the same relation given in Eq. (\ref{scale}) as the optimal Stokes efficiency.}
\end{figure}

\section{Discussion and conclusion}

In this paper, we search for general principles of designing key microscopic motor properties, specifically the interaction potential $V(\theta)$ and the stepping rate function $k_+(\theta)$,  in order to optimize the macroscopic motor performance characterized by its power output and efficiency. Different from previous work, we have taken into account realistic biophysical and biochemical constraints on the shape of $V$ and the reversibility of the mechanochemical cycles ($\varepsilon <1$, $\kappa<1$) in our investigation. We have studied the detailed dynamics and energetics of the high-performing bacterial flagellar motor in comparison with quantitative experimental data in order to test our general theory, which should be applicable to other molecular motors as well. 
In the following, we discuss our main general findings and their applications to the BFM together with related work from other groups:

(1) A motor's power output depends on its torque(or force)-speed dependence. According to our theoretical analysis and simulations, a gating mechanism that allows the ion-assisted stator conformation to occur in a narrow window of relative positions between the stator and the rotor can lead to the observed concave torque-speed curve in CCW BFM. The concavity of the torque-speed curve increases with the gating strength. As a result, the maximum power output, which occurs at an intermediate load level near the knee of the torque-speed curve, increases with the gating strength. In general, a strong gating regime is a key design feature for $k_+(\theta)$ in order to generate maximum power in a wide range of physiologically relevant loads. Our results also provide a plausible explanation for the almost linear torque-speed curve for the CW state \cite{Yuan2010}: the gating strength may be weaker in the CW state. The molecular mechanism for gating is unclear, it requires more structural and biochemical studies of the rotor-stator interaction and its effect on regulating ion translocation.


(2) 
The conventional definition of motor efficiency ($\Lambda_0$) \cite{Wang2002} implicitly assumes tight-coupling, i.e., all backward steps regain chemical energy by pumping out ions in the case of BFM or synthesizing ATP in the case of linear motors. 
In reality, there may be only a fraction $\kappa <1$ of back steps that regain energy. In the case of the linear motor kinesin, experiments show that ATP hydrolysis rate is finite even at stall when there are equal number of forward and backward steps and some backward steps can even cost energy \cite{Carter2005,Lipowsky2007}.  Here, we show that efficiency peaks at a finite speed and the maximum efficiency is less than $1$ as long as there is a finite spontaneous stepping probability, i.e., $\kappa<1$. 

In a recent paper\cite{Boschert2015}, Boschert et al. proposed a loose coupling model to explain the less-than-two ions translocation per step in the bacterial flagella motor observed in \cite{Lo2013}. The model was based purely on the conformational changes of the stator without considering the motor's actual physical rotation. It was assumed that the motor can generate a constant torque (or perform work) with either one or two ions bound, but the work done is the same regardless of whether one or two ions passes the membrane. 
The case of torque generation by two ion translocations can be considered as two forward steps followed by a "wasteful" back step. The assumed finite probability of a power stroke by the stator with two ions bound is consistent with an effective $\kappa<1$ in our model.

The existence of back steps in BFM is strongly suggested \cite{Meacci2011} by the observed continuity of torque when motors are forced to rotate with a small negative speed \cite{Berry1997}. Otherwise, the motor would show a barrier in its torque-speed curve near stall, which was not observed. However, it is not clear whether all back steps pump out ions. We suspect the spontaneous back steps are not negligible, i.e., $\kappa<1$. Future experiments that directly measure ion translocation, specially during forced slow back rotations \cite{Berry1997}, are needed to test this hypothesis.

(3) We have derived two thermodynamics laws for the nonequilibrium motor. By using these laws for BFM, we showed that the maximum torque at stall should be strictly less than $G_0/\theta_0$ for any biologically realistic form of $ V(\theta)$, including the electro-steric potential proposed recently by Mandadapu et al. \cite{Mandadapu2015}. The difference $G_0/\theta_0-\tau_{max} \approx \sigma_{\tau}/\tau_{max}$ is mostly due to torque and speed fluctuations at high loads. 

In general, the design of the interaction potential $V$ to optimize the maximum torque (force) and the motor efficiency is dictated by the tradeoff of two opposing effects of the energy gap $E_g$. For a given energy budget $G_0=\Delta V+E_g$, a steep $V(\theta)$ leads to a large $\Delta V$, which increases torque, but at the same time a finite positive $E_g$ is also needed to suppress backward steps, which have the adverse effects of slowing down the motor and wasting energy. 
As a result of this tradeoff, we obtain a general limit for the optimal efficiency $\Lambda^*$: $1-\Lambda^* \propto \ln(E_0)/E_0$. A high efficiency (Eq. (\ref{scale})) can be achieved at the choice of an optimum energy gap $E_g^*(>0)$ that depends logarithmically on $E_0$ for large $E_0$. 


Our model can naturally explain the recent experiments \cite{Lo2013} reporting $\tau_{max}$ being around $0.71 E_0/\theta_0$. 
From our study, this experimental observation indicates an energy gap 
$E_g/E_0\approx 0.29$, which is close to the optimal values of $E_g$ resulting from maximizing the power or the efficiency (see Fig.\,4A\&B). It remains an interesting open question as to whether the motor has evolved to optimize its performance measured by power output, efficiency, or a combination of the two under physiological constraints. The gerenal model framework should be useful in understanding energetics of other molecular motors. Our results here may also provide guidance in designing more efficient and powerful synthetic motors \cite{Astumian2015}.  

\begin{acknowledgments}
We thank Dr. B. Hu for discussions in early stage of the work. We aslo thank Dr. C-J Lo for sharing data from \cite{Lo2013} and Drs. Howard Berg and Joe Howard for useful discussions. This work is supported by the National Institutes of Health Grant GM081747 (YT).
\end{acknowledgments}

\appendix

\section{Detailed derivation of the Fokker-Planck equation for the minimal motor model}
There are two processes in motor dynamics, a continuous noisy mechanical motion and discrete stochastic chemical transitions, which can be described by a Langevin equation and the chemical transition rates, respectively:   -- that can be described by can be described by
\begin{eqnarray}
&&\xi \frac{d\theta_R}{dt} = -V'(\theta_R-\theta_S)+\eta(t), \label{lang}\\
&&\mathrm{Prob}(\theta_S\rightarrow \theta_S \pm \Delta \theta) = k_{\pm}(\theta)dt, \label{step}
\end{eqnarray}
where $\eta$ represents the white thermal noise: $<\eta(t)\eta(t')>=2\xi k_BT\delta(t-t')$ ($k_BT$ is the thermal energy set to $1$) and $\Delta\theta=\theta_0$ is the step size of chemical transitions. A stator stepping event results in a shift of the interaction potential in the direction of the motor rotation by an angle $\theta_0$ and the subsequent motor motion is governed by this new potential until the next stepping event occurs. The stepping rates have a periodicity of $2\theta_0$, i.e., $k_{\pm}(\theta)=k_{\pm}(\theta+2\theta_0)$. For physical motion Eq. (\ref{lang}), we have assumed for simplicity that the rotor and the external load move in unison and denoted their total drag coefficient by $\xi$. .

Although only two energy landscapes are plotted in Fig.\,1A in the main text, the model contains $52$ such landscapes. By symmetry and periodicity, once the motor steps forward to the third landscape (which is not shown), the process effectively repeats itself as starting from the first landscape (shown in Fig.\,1A). Therefore, this model is equivalent to a particle moving along only two energy landscapes, $V_1(\theta)$ and $V_2(\theta)$, which have the same shape and only differ by a half-period shift: $V_1(\theta)=V(\theta)$, $V_2(\theta)=V(\theta+\theta_0)$. 

\begin{figure}
\centering
\includegraphics[width=.5\textwidth,trim={1.5cm 4.0cm 1.5cm 4.5cm},clip]{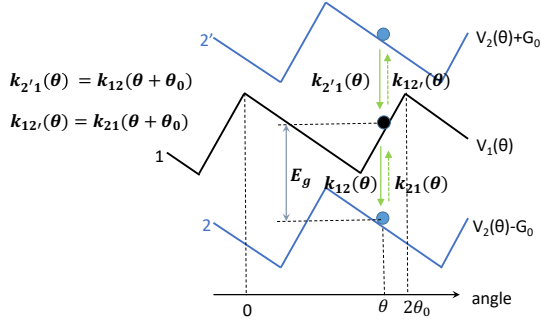}
\caption{ Illustration of transitions between states in different energy landscapes shifted by the IMF ($G_0$). The black line represents the state $1$ with potential $V_1(\theta)=V(\theta)$. The blue lines represent the two adjacent states ($2'$ and $2$) with potentials $V_2(\theta)=V(\theta+\theta_0)$ shifted by $G_0$ and $-G_0$ for state $2'$ and state $2$ respectively, $\theta_0$ is the half period. The green arrowed lines represent the transitions between the states with the transition rates labeled (see Appendix A for details). Due to symmetry, the transition rates ($k_{2'1}$ and $k_{12'}$) between states $2'$ and $1$ are the same as the transition rates ($k_{12}$ and $k_{21}$) between states $1$ and $2$ with the angle $\theta$ shifted by $\theta_0$. }
\label{FigS0}
\end{figure}

The system can be described by two coupled Fokker-Planck equations governing the probabilities, $P_1(\theta,t)$ and $P_2(\theta,t)$, of the particle in each of the two energy landscapes:
\begin{eqnarray}
\frac{\partial P_1(\theta,t)}{\partial t}&=&\xi^{-1}\frac{\partial}{\partial \theta}[V_1'(\theta)P_1+P'_1]+\Delta j_s(\theta) ,\label{FP_1} \\
\frac{\partial P_2(\theta,t)}{\partial t}&=&\xi^{-1}\frac{\partial}{\partial \theta}[V_2'(\theta)P_2+P'_2]-\Delta j_s(\theta),
\label{FP_2}
\end{eqnarray}
where the net flux due to jumping transitions between $V_1$ and $V_2$ is given by $\Delta j_s(\theta)$, which can be expressed as:
\begin{equation}
\Delta j_s(\theta)=[k_{21}(\theta)+k_{2'1}(\theta)]P_2(\theta)-[k_{12}(\theta)+k_{12'}(\theta)]P_1(\theta),
\label{js}
\end{equation}
where $k_{12}(\theta)$ represents the transition rate from the first energy landscape $V_1$, called state $1$, to the second landscape $V_2$ downshifted by the effective driving energy $G_0$ (state $2$) and $k_{21}(\theta)$ is the corresponding reverse transition rate. Included in $\Delta j_s$ are also transitions between $V_1$ and the previous $V_2$ energy landscape shifted up by $G_0$ (called state $2'$), as illustrated in Fig.\,7. Due to symmetry between the two states ($1$ and $2$), the transition rates between state $2'$ and state $1$ are the same as those between state $1$ and state $2$, only shifted by $\theta_0$: $$k_{2'1}(\theta)=k_{12}(\theta+\theta_0),\;\;\; k_{12'}(\theta)=k_{21}(\theta+\theta_0),$$ as shown in Fig.\,7. All these stepping transitions are included in the expression for $\Delta j_s$ above.  

All these functions, including $V$, $P_1$, $P_2$, $k_{12}$, and $k_{21}$, are periodic functions with the full period $2\theta_0$. By symmetry, we also have $P_1(\theta)=P_2(\theta+\theta_0)$. Using these relationships and defining $P_1(\theta,t)\equiv P(\theta,t)$, we have $P_2(\theta,t)=P(\theta+\theta_0,t)$, and the two coupled Fokker-Planck equations, Eq.(\ref{FP_1}-\ref{FP_2}), can be combined into one equation for $P(\theta,t)$ given as Eq. (\ref{FPE}) in the main text. For convenience of formulating a single Fokker-Planck equation, we use $k_{\pm}$ instead of $k_{12}$ and $k_{21}$:
$
k_{12}(\theta)\equiv k_+(\theta),\;\;\; k_{21}(\theta)\equiv k_-(\theta-\theta_0).
$


A good design of $k_{+}(\theta)$ is to allow energy-assisted forward steps to occur only in the half-period region $[\theta_0,2\theta_0)$ so that the stator can ``jump" onto the next energy landscape to continue generating positive torque. Therefore, it is favorable to have nonzero $k_{+}(\theta)\neq0$ only for $\theta\in[\theta_0,2\theta_0)$. In this paper, we assume $k_+(\theta)=0$ for $\theta\in[0,\theta_0)$. Correspondingly, Eq. (\ref{db1}) requires that $k_-(\theta)=0$ for $\theta\in[\theta_0, 2\theta_0)$. Therefore, we can express $k_{12}$ and $k_{21}$ in terms of $k_{\pm}$:
\begin{equation}
k_{12}(\theta)=\begin{cases}
k_+(\theta) \  & \text{for} \  \theta\in[\theta_0,2\theta_0), \\
0  \ & \text{for} \  \theta\in[0,\theta_0).
\end{cases}
\label{k12}
\end{equation}
\begin{equation}
 k_{21}(\theta)=\begin{cases}
k_-(\theta-\theta_0) \  & \text{for} \  \theta\in[\theta_0,2\theta_0), \\
0  \  & \text{for} \  \theta\in[0,\theta_0).
\end{cases}
\label{k21}
\end{equation}
By plugging the above expressions for $k_{12}$ and $k_{21}$ into Eq. (\ref{js}), we have the expression for $\Delta j_s(\theta)$ as shown in Eq. (\ref{ss}) in the main text.


\section{Details of the model and parameters}

An energy barrier $V_B$ near the peak of $V(\theta)$ is added to prevent slipping between two adjacent FliG's without stepping. A linear form is used: $V_B(\theta)= H(\theta_{B}-\theta)/\theta_B$ for $\theta\in[0, \theta_{B}]$; $V_B(\theta)=H (\theta-\theta_0+\theta_{B})/\theta_{B}$ for $\theta\in[2\theta_0-\theta_{B},2\theta_0)$; $0$, otherwise. The barrier height is $H\gg k_B T$, and its width is $\theta_B\ll\theta_0$. 


The standard parameters used in this paper are based on previous modeling studies and by fitting our model to available experimental data: $\theta_0=\pi/26$, $\theta_{B}=0.005\theta_0$, $\theta_g=0.05\theta_0$, $k_a=0$, $k_b=1\times 10^4s^{-1}$, $k_g=5\times10^5s^{-1}$, $H=50k_BT$, $\xi=10^{-3}- 10^3 (pN\cdot nm \cdot rad^{-1} \cdot s)$, $E_0=10 k_BT$, $\varepsilon\in (0,1)$, $\kappa\in(0,1]$, $k_BT=4.11(pN\cdot nm)$ for room temperature. Unless specifically mentioned, we used $\varepsilon=0.5$ and $\kappa=0.5$ in the main text. The units of the parameters are omitted in the main text of the paper, they are the same as given here. 

\section{The dependence of $P_s(\theta)$ on $E_g$} The steady state distribution $P_s(\theta)$ depends on the energy gap $E_g=G_0-V_d/(1+\varepsilon)=G_0-\tau_+\theta_0$. As explained in the main text and shown in Fig.\,8, when $E_g$ decreases the probability in the negative torque regime $P_-\equiv \int_{\theta_m}^{2\theta_0} P_s(\theta)d\theta$ increases and thus the probability in the positive torque regime $P_+=1-P_-$ decreases. Together with the fact that $\tau_+$ increases with a decreasing $E_g$, this explains the peak in the maximum torque $\tau_m$ seen in Fig.\,3A. 

\begin{figure}[!ht]
\begin{center}
\includegraphics[width=.5\textwidth,trim={5.0cm 8.5cm 5.0cm 1.5cm},clip]{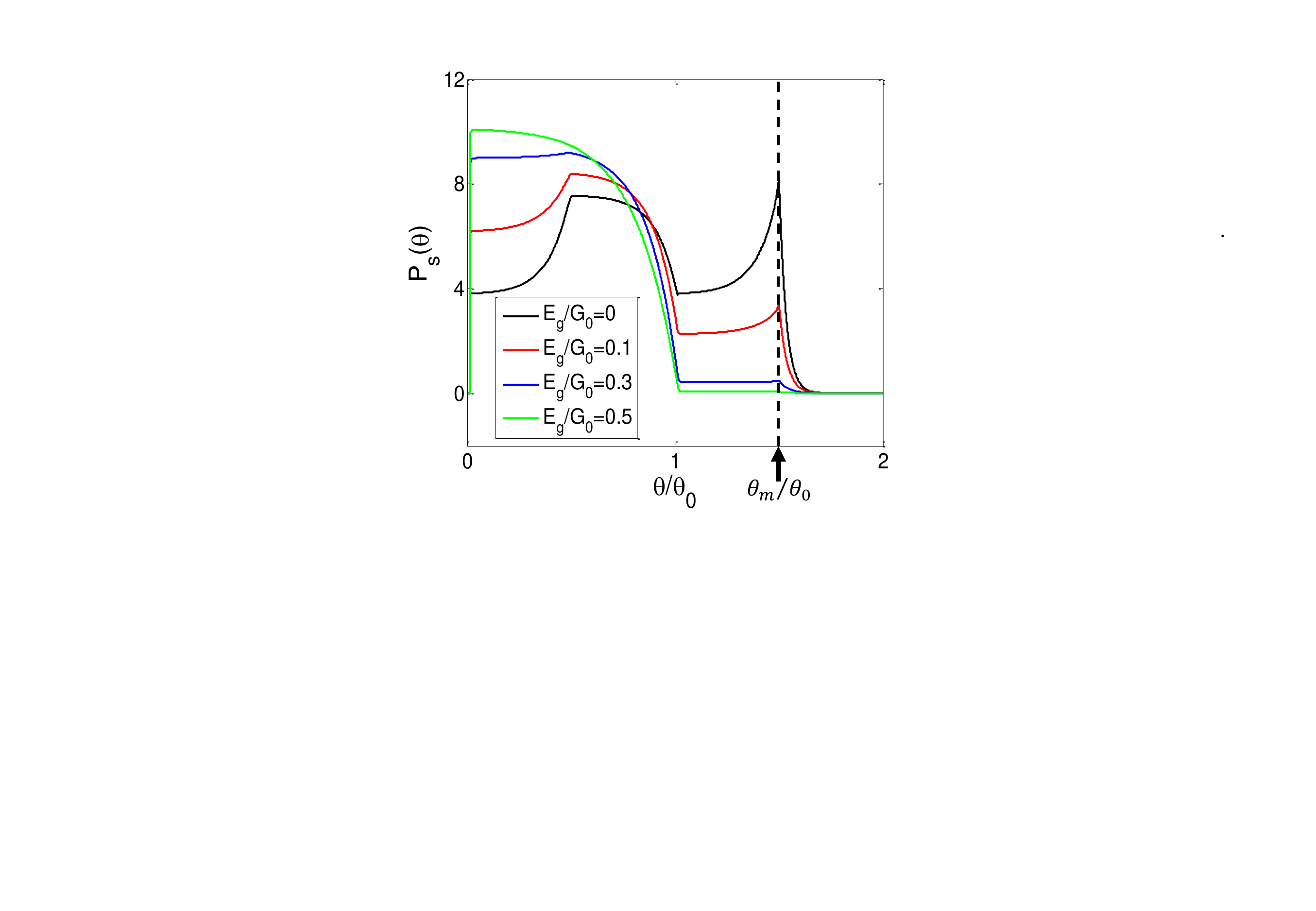}
\end{center}
\caption{
The distribution $P_s(\theta)$ for different values of $E_g$ (by varying $V_d$ and fixing $\varepsilon=0.5$). The functions $V(\theta)$ and $k_+(\theta)$ are the same as shown in Fig. 1C in the main text. The probability in the negative torque regime $P_-\equiv \int_{\theta_m}^{2\theta_0} P_s(\theta)d\theta$ decreases with increasing $E_g$ and thus the probability in the positive toque regimes $P_+=1-P_-$ decreases with decreasing $E_g$.  
}
\label{FigS2}
\end{figure}

\section{Results with quadratic $V(\theta)$}

Besides the V-shaped piecewise linear form of $V(\theta)$ used in the main text,  we have also used other form of $V(\theta)$, such as the quadratic form given as:
\begin{eqnarray}
V(\theta)&=&\frac{V_d  (\theta_m-\theta)^2}{\theta_m^2}, \;\;\; 0<\theta\le\theta_m,\nonumber \\ &=& \frac{V_d (\theta-\theta_m)^2 }{(2\theta_0-\theta_m)^2}, \;\;\; \theta_m<\theta\le 2\theta_0,
\label{qua}
\end{eqnarray}
which is shown in Fig.\,9A. The overall shape of the quadratic potential is given by its depth $V_d$ and it off-centered minimum location  $(1+\varepsilon)\theta_0$. For such a quadratic potential we repeated what we did in the main text with the energy gap defined as : $E_g=G_0-V_d(1+2\varepsilon)/(1+\varepsilon)^2$. The results on the maximum torque versus $1-E_g/G_0$, the maximum efficiency versus $ln(E_0)/E_0$, and the optimal $E^*_g$ versus $\ln E_0$ are shown in Fig.\,9B\&C\&D, respectively, which are similar to the results shown in the main text with the piece-wise linear potential. 

\begin{figure}[!ht]
\centering
\includegraphics[width=.5\textwidth]{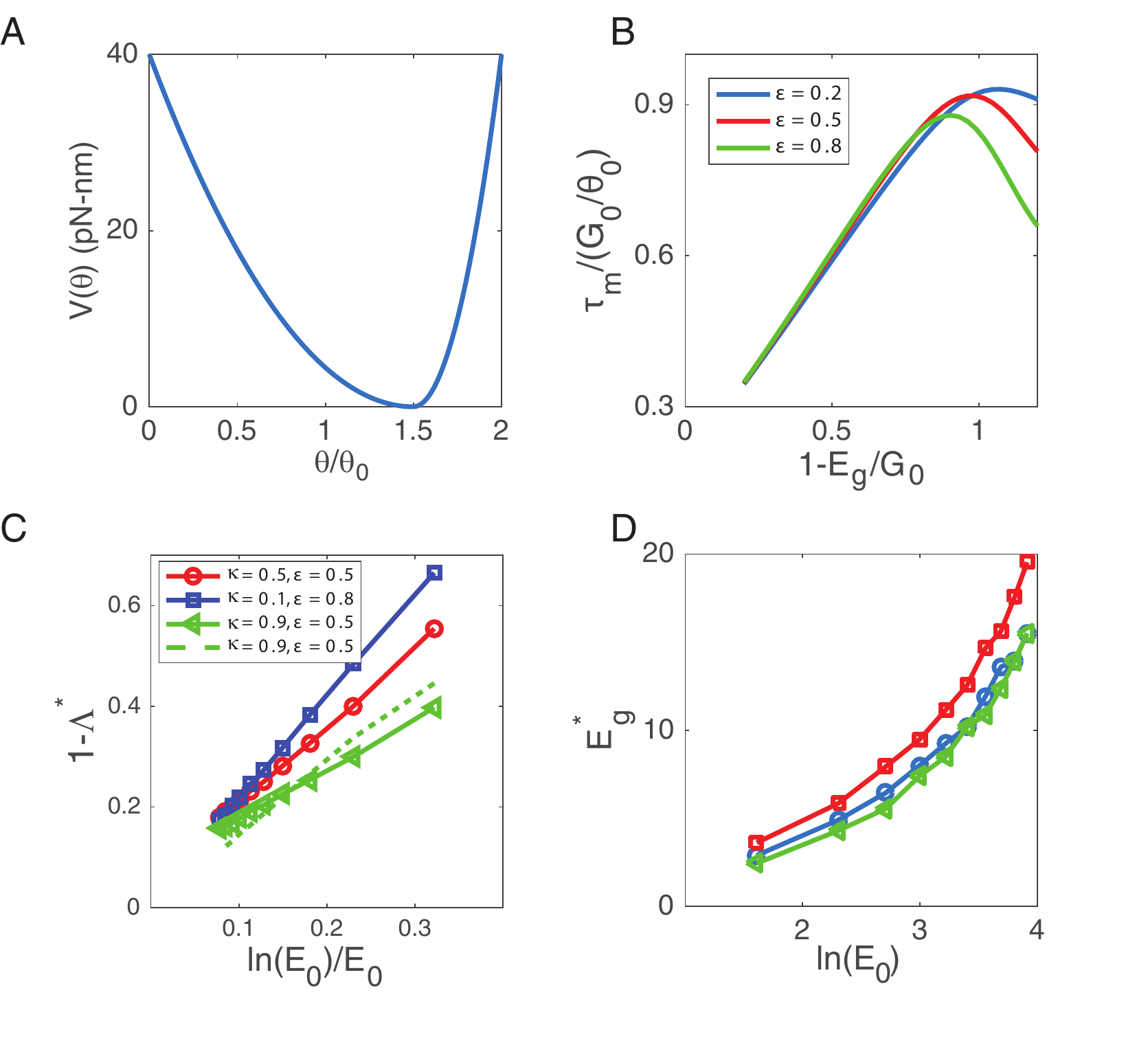}
\caption{
The results for quadratic potentials. (A) The shape of a quadratic potential, expression given in Eq. (\ref{qua}), in which the minimum location at $\theta_m=(1+\varepsilon)\theta_0$ and the depth of the potential $V_d$ are varied to optimize motor performance. (B) The maximum torque versus $1-E_g/G_0=V_d(1+2\varepsilon)/[G_0(1+\varepsilon)^2]$. (C) $1-\Lambda^*$  versus $ln(E_0)/E_0$, where $\Lambda^*$ is the maximum efficiency. (D) The optimal $E^*_g$ (for achieving the maximum efficiency)  versus $\ln E_0$.
}
\end{figure}

\end{document}